\documentclass[12pt]{article}
\usepackage{times}
\usepackage{geometry}
\usepackage{enumitem,amssymb}
\usepackage{ragged2e}
\newlist{thematic}{itemize}{8}
\setlist[thematic]{label=$\square$}
\usepackage{pifont}

\usepackage{graphicx}
\usepackage{wrapfig}
\usepackage{amsmath,amssymb,color}

\newcommand{\n}{\noindent}

\newcommand{\Amherst}{University of Massachusetts, Amherst, MA 01003 USA}
\newcommand{\ANLHEP}{HEP Division, Argonne National Laboratory, Lemont, IL 60439, USA}
\newcommand{\APC}{Laboratoire Astroparticule et Cosmologie (APC), CNRS/IN2P3, Universit\'e Paris Diderot, 10, rue Alice Domon et Léonie Duquet, 75205 Paris Cedex 13, France}

\newcommand{\AUCK}{University of Auckland, Private Bag 92019 Auckland, New Zealand}
\newcommand{\BenGurion}{Department of Physics, Ben-Gurion University, Be'er Sheva 84105, Israel}
\newcommand{\BNL}{Brookhaven National Laboratory, Upton, NY 11973}
\newcommand{\Brown}{Brown University, Providence, RI 02912}

\newcommand{\BU}{Boston University, Boston, MA 02215}
\newcommand{\Buffalo}{Department of Physics, University at Buffalo, SUNY Buffalo, NY 14260 USA}
\newcommand{\Caltech}{California Institute of Technology, Pasadena, CA 91125}

\newcommand{\Cavendish}{Astrophysics Group, Cavendish Laboratory, J.J.Thomson Avenue, Cambridge, CB3 0HE, UK}
\newcommand{\CCA}{Center for Computational Astrophysics, 162 5th Ave, 10010, New York, NY, USA}

\newcommand{\CEADAP}{D\'epartement d?Astrophysique, CEA Saclay DSM/Irfu, 91191 Gif-sur-Yvette, France}

\newcommand{\CfA}{Harvard-Smithsonian Center for Astrophysics, MA 02138}

\newcommand{\Cincinnati}{University of Cincinnati, Cincinnati, OH 45221}
\newcommand{\CITA}{Canadian Institute for Theoretical Astrophysics, University of Toronto, Toronto, ON M5S 3H8, Canada}

\newcommand{\Columbia}{Columbia University, New York, NY 10027}
\newcommand{\Cornell}{Cornell University, Ithaca, NY 14853}

\newcommand{\CWRU}{Case Western Reserve University, Cleveland, OH 44106}

\newcommand{\damtp}{DAMTP, Centre for Mathematical Sciences, Wilberforce Road, Cambridge, UK, CB3 0WA}

\newcommand{\DFI}{Departamento de F\'isica, FCFM, Universidad de Chile, Blanco Encalada 2008, Santiago, Chile}

\newcommand{\dunlap}{Dunlap Institute for Astronomy and Astrophysics, University of Toronto, ON, M5S3H4}

\newcommand{\FNAL}{Fermi National Accelerator Laboratory, Batavia, IL 60510}

\newcommand{\GRAPPA}{GRAPPA Institute, University of Amsterdam, Science Park 904, 1098 XH Amsterdam, The Netherlands}
\newcommand{\GSFC}{Goddard Space Flight Center, Greenbelt, MD 20771 USA}

\newcommand{\HarvardPhys}{Department of Physics, Harvard University, Cambridge, MA 02138, USA}
\newcommand{\Haverford}{Haverford College, 370 Lancaster Ave, Haverford PA, 19041, USA}

\newcommand{\HKUST}{The Hong Kong University of Science and Technology, Hong Kong SAR, China}

\newcommand{\IAP}{Institut d'Astrophysique de Paris (IAP), CNRS \& Sorbonne University, Paris, France}
\newcommand{\IAS}{Institute for Advanced Study, Princeton, NJ 08540}
\newcommand{\IBS}{Institute for Basic Science (IBS), Daejeon 34051, Korea}

\newcommand{\ICE}{Institute of Space Sciences (ICE, CSIC), Campus UAB, Carrer de Can Magrans, s/n, 08193 Barcelona, Spain}

\newcommand{\ICTP}{International Centre for Theoretical Physics, Strada Costiera, 11, I-34151 Trieste, Italy}

\newcommand{\IFT}{Instituto de Fisica Teorica UAM/CSIC, Universidad Autonoma de Madrid, 28049 Madrid, Spain}

\newcommand{\INFNFE}{Istituto Nazionale di Fisica Nucleare, Sezione di Ferrara, 40122, Italy }

\newcommand{\INFNRM}{Istituto Nazionale di Fisica Nucleare, Sezione di Roma, 00185 Roma, Italy}

\newcommand{\ioa}{Institute of Astronomy, University of Cambridge,Cambridge CB3 0HA, UK}

\newcommand{\IPMU}{Kavli Insitute for the Physics and Mathematics of the Universe (WPI), University of Tokyo, 277-8583 Kashiwa , Japan}

\newcommand{\ITFA}{Institute for Theoretical Physics, University of Amsterdam, Science Park 904, 1098 XH Amsterdam, The Netherlands}

\newcommand{\JHU}{Johns Hopkins University, Baltimore, MD 21218}

\newcommand{\JPL}{Jet Propulsion Laboratory, California Institute of Technology, Pasadena, CA, USA}
\newcommand{\kagawa}{Department of General Education, National Institute of Technology, Kagawa College, 355Chokushi-cho, Takamatsu, Kagawa 761-8058, Japan}

\newcommand{\kavli}{Kavli Institute for Cosmology, Cambridge, UK, CB3 0HA}
\newcommand{\kenyon}{Department of Physics, Kenyon College, 201 N College Rd, Gambier, OH 43022}
\newcommand{\KIAS}{School of Physics, Korea Institute for Advanced Study, 85 Hoegiro, Dongdaemun-gu, Seoul 130-722, Korea}
\newcommand{\KICP}{Kavli Institute for Cosmological Physics, Chicago, IL 60637}
\newcommand{\KIPAC}{Kavli Institute for Particle Astrophysics and Cosmology, Stanford 94305}
\newcommand{\KINGS}{King's College London, WC2R 2LS London, United Kingdom}

\newcommand{\LBL}{Lawrence Berkeley National Laboratory, Berkeley, CA 94720}

\newcommand{\LLNL}{Lawrence Livermore National Laboratory, Livermore, CA, 94550}

\newcommand{\Melbourne}{School of Physics, The University of Melbourne, Parkville, VIC 3010, Australia}

\newcommand{\MIT}{Massachusetts Institute of Technology, Cambridge, MA 02139}

\newcommand{\NAOC}{National Astronomical Observatories, Chinese Academy of Sciences, PR China}
\newcommand{\NBI}{The Niels Bohr Institute \& Discovery Center, Blegdamsvej 17, DK-2100 Copenhagen, Denmark}

\newcommand{\NSW}{University of New South Wales, Sydney NSW 2052, Australia}

\newcommand{\OSU}{The Ohio State University, Columbus, OH 43212}

\newcommand{\OskarKlein}{Oskar Klein Centre for Cosmoparticle Physics, Stockholm University, AlbaNova, Stockholm SE-106 91, Sweden}
\newcommand{\Oxford}{The University of Oxford, Oxford OX1 3RH, UK}

\newcommand{\ParisSud}{Universit\'{e} Paris-Sud, LAL, UMR 8607, F-91898 Orsay Cedex, France \& CNRS/IN2P3, F-91405 Orsay, France}
\newcommand{\PI}{Perimeter Institute, Waterloo, Ontario N2L 2Y5, Canada}

\newcommand{\Port}{Institute of Cosmology \& Gravitation, University of Portsmouth, Dennis Sciama Building, Burnaby Road, Portsmouth PO1 3FX, UK}
\newcommand{\Princeton}{Princeton University, Princeton, NJ 08544}

\newcommand{\Rice}{Department of Physics \& Astronomy, Rice University, Houston, Texas 77005, USA}

\newcommand{\RomaS}{Dipartimento di Fisica, Universit\`{a} La Sapienza, P. le A. Moro 2, Roma, Italy}

\newcommand{\Sejong}{Department of Physics and Astronomy, Sejong University, Seoul, 143-747, Korea}

\newcommand{\SHAO}{Shanghai Astronomical Observatory (SHAO), Nandan Road 80, Shanghai 200030, China}

\newcommand{\SimonFraser}{Department of Physics, Simon Fraser University, Burnaby, British Columbia, Canada V5A 1S6}
\newcommand{\SLAC}{SLAC National Accelerator Laboratory, Menlo Park, CA 94025}
\newcommand{\SMU}{Southern Methodist University, Dallas, TX 75275}

\newcommand{\Stanford}{Stanford University, Stanford, CA 94305}
\newcommand{\StonyBrook}{Stony Brook University, Stony Brook, NY 11794}

\newcommand{\SussexAstronomy}{Astronomy Centre, School of Mathematical and Physical Sciences, University of Sussex, Brighton BN1 9QH, United Kingdom}
\newcommand{\Syracuse}{Syracuse University, Syracuse, NY 13244}

\newcommand{\UAM}{Universidad Aut\'onoma de Madrid, 28049, Madrid, Spain}
\newcommand{\UBC}{University of British Columbia, Vancouver, BC V6T 1Z1, Canada}

\newcommand{\UCBP}{Department of Physics, University of California Berkeley, Berkeley, CA 94720, USA}

\newcommand{\UCD}{University of California at Davis, Davis, CA 95616}
\newcommand{\UChicago}{University of Chicago, Chicago, IL 60637}
\newcommand{\UCI}{University of California, Irvine, CA 92697}
\newcommand{\UCLA}{University of California at Los Angeles, Los Angeles,  CA 90095}
\newcommand{\UCL}{University College London, WC1E 6BT London, United Kingdom}

\newcommand{\UCSD}{University of California San Diego, La Jolla, CA 92093}
\newcommand{\UFL}{University of Florida, Gainesville, FL 32611}

\newcommand{\UGTO}{Divisi\'on de Ciencias e Ingenier\'ias, Universidad de Guanajuato, Le\'on 37150, M\'exico}

\newcommand{\UMich}{University of Michigan, Ann Arbor, MI 48109}
\newcommand{\UMN}{University of Minnesota, Minneapolis, MN 55455}

\newcommand{\UNIPD}{Dipartimento di Fisica e Astronomia ``G. Galilei'',Universit\`a degli Studi di Padova, via Marzolo 8, I-35131, Padova, Italy}
\newcommand{\UNM}{University of New Mexico, Albuquerque, NM 87131}

\newcommand{\UoM}{Jodrell Bank Center for Astrophysics, School of Physics and Astronomy, University of Manchester, Oxford Road, Manchester, M13 9PL, UK}
\newcommand{\UPenn}{Department of Physics and Astronomy, University of Pennsylvania, Philadelphia, Pennsylvania 19104, USA}

\newcommand{\UrbanaC}{Department of Physics, University of Illinois at Urbana-Champaign, Urbana, Illinois 61801, USA}

\newcommand{\UTD}{University of Texas at Dallas, Texas 75080}

\newcommand{\UWMadison}{Department of Physics, University of Wisconsin - Madison, Madison, WI 53706}
\newcommand{\UW}{University of Washington, Seattle 98195}
\newcommand{\UWC}{Department of Physics \& Astronomy, University of the Western Cape, Cape Town 7535, South Africa}

\newcommand{\VSI}{Van Swinderen Institute for Particle Physics and Gravity, University of Groningen, Nijenborgh 4, 9747~AG~Groningen, The~Netherlands}

\newcommand{\Yale}{Department of Physics, Yale University, New Haven, CT 06520}
\newcommand{\YorkU}{Department of Physics and Astronomy, York University, Toronto, Ontario M3J 1P3, Canada}

\begin{document}
 \pagenumbering{gobble}
{\raggedright
\huge
Astro2020 Science White Paper \linebreak

Probing the origin of our Universe through cosmic microwave background constraints on gravitational waves \linebreak
\normalsize

\noindent \textbf{Thematic Areas:} \hspace*{60pt} $\square$ Planetary Systems \hspace*{10pt} $\square$ Star and Planet Formation \hspace*{20pt}\linebreak
$\square$ Formation and Evolution of Compact Objects \hspace*{31pt} X Cosmology and Fundamental Physics \linebreak
  $\square$  Stars and Stellar Evolution \hspace*{1pt} $\square$ Resolved Stellar Populations and their Environments \hspace*{40pt} \linebreak
  $\square$    Galaxy Evolution   \hspace*{45pt} $\square$             Multi-Messenger Astronomy and Astrophysics \hspace*{65pt} \linebreak
  
\textbf{Principal Author:}

Name: Sarah Shandera	
 \linebreak						
Institution:  The Pennsylvania State University
 \linebreak
Email: ses47@psu.edu
 \linebreak
Phone:  (814)863-9595
 \linebreak

\textbf{Co-authors:} Peter Adshead$^{1}$, Mustafa Amin$^{2}$, Emanuela Dimastrogiovanni$^{3}$, Cora Dvorkin$^{4}$, Richard Easther$^{5}$, Matteo Fasiello$^{6}$, Raphael Flauger$^{7}$, John T. Giblin,~Jr$^{8}$,  Shaul Hanany$^{9}$, Lloyd Knox$^{10}$, Eugene Lim$^{11}$, Liam McAllister$^{12}$, Joel Meyers$^{13}$, Marco Peloso$^{14}$,  Graca Rocha$^{15}$, Maresuke Shiraishi$^{16}$, Lorenzo Sorbo$^{17}$, Scott Watson$^{18}$
  \linebreak

\n{\it $^{1}$\UrbanaC \\\n $^{2}$ \Rice \\\n  $^{3}$ \NSW \\\n $^{3}$ \NSW \\\n $^{4}$ \HarvardPhys \\\n $^{5}$ \AUCK \\\n $^{6}$ \Port \\\n  $^{7}$ \UCSD \\\n $^{8}$ \kenyon \\\n  $^{9}$ \UMN \\\n $^{10}$ \UCD \\\n  $^{11}$ \KINGS \\\n  $^{12}$ \Cornell \\\n  $^{13}$ \SMU \\\n $^{14}$ \UNIPD \\\n $^{15}$ \JPL \\\n $^{16}$ \kagawa \\\n $^{17}$ \Amherst \\\n $^{18}$ \Syracuse \\\n

}

\pagebreak
 }
\n\textbf{Endorsers:} Zeeshan Ahmed$^{1}$, 
David Alonso$^{2}$, 
Robert Armstrong$^{3}$, 
Mario Ballardini$^{4}$, 
Darcy Barron$^{5}$, 
Nicholas Battaglia$^{6}$, 
Daniel Baumann$^{7,8}$, 
Charles Bennett$^{9}$, 
Bradford Benson$^{10,11}$, 
Florian Beutler$^{12}$, 
Colin Bischoff$^{13}$, 
Lindsey Bleem$^{14,11}$, 
J. Richard Bond$^{15}$, 
Julian Borrill$^{16}$, 
Cliff Burgess$^{17}$, 
Victor Buza$^{18}$, 
Christian T. Byrnes$^{19}$, 
Erminia Calabrese$^{}$, 
John E.\ Carlstrom$^{20,11,14}$, 
Sean Carroll$^{21}$, 
Anthony Challinor$^{22,23,24}$, 
Xingang Chen$^{25}$, 
Asantha Cooray$^{26}$, 
Thomas Crawford$^{20,11}$, 
Francis-Yan Cyr-Racine$^{18,5}$, 
Guido D'Amico$^{27}$, 
Paolo de Bernardis$^{28,29}$, 
Jacques Delabrouille$^{30,31}$, 
Marcel~Demarteau$^{14}$, 
Olivier Dor\'e$^{32}$, 
Duan Yutong$^{33}$, 
Joanna Dunkley$^{34}$, 
Jeffrey Filippini$^{35}$, 
Simon Foreman$^{15}$, 
Pablo Fosalba$^{36}$, 
Aur\'elien A.~Fraisse$^{34}$, 
Fran\c{c}ois R. Bouchet$^{37}$, 
Juan Garc\'ia-Bellido$^{38}$, 
Juan Garc\'ia-Bellido$^{39}$, 
Martina Gerbino$^{14}$, 
Vera Gluscevic$^{40}$, 
Garrett Goon$^{23}$, 
Krzysztof M. G\'orski$^{41}$, 
Daniel Grin$^{42}$, 
Jon E. Gudmundsson$^{43}$, 
Nikhel Gupta$^{44}$, 
Shaul Hanany$^{45}$, 
Will Handley$^{24,46}$, 
J.~Colin~Hill$^{47,48}$, 
Christopher M. Hirata$^{49}$, 
Gilbert Holder$^{35}$, 
Dragan Huterer$^{50}$, 
Mustapha Ishak$^{51}$, 
Bradley Johnson$^{52}$, 
Matthew C. Johnson$^{17,53}$, 
William C. Jones$^{34}$, 
Kenji Kadota$^{54}$, 
Marc Kamionkowski$^{9}$, 
Kirit S. Karkare$^{20,11}$, 
Nobuhiko Katayama$^{55}$, 
William~H.~Kinney$^{56}$, 
Theodore Kisner$^{16}$, 
Lloyd Knox$^{57}$, 
Savvas M. Koushiappas$^{58}$, 
Ely D.~Kovetz$^{59}$, 
Kazuya Koyama$^{12}$, 
Massimiliano Lattanzi$^{60}$, 
Hayden Lee$^{18}$, 
Marilena Loverde$^{61}$, 
Silvia Masi$^{28,29}$, 
Kiyoshi Masui$^{62}$, 
Liam McAllister$^{6}$, 
Jeff McMahon$^{50}$, 
Matthew McQuinn$^{63}$, 
P.~Daniel Meerburg$^{24,23,64}$, 
P.~Daniel Meerburg$^{24,64,23}$, 
Joel Meyers$^{65}$, 
Mehrdad Mirbabayi$^{66}$, 
Pavel Motloch$^{15}$, 
Suvodip Mukherjee$^{37}$, 
Julian B.~Mu\~noz$^{18}$, 
Johanna Nagy$^{67}$, 
Pavel Naselsky$^{68}$, 
Federico Nati$^{}$, 
Laura Newburgh$^{69}$, 
Michael D. Niemack$^{6}$, 
Gustavo Niz$^{70}$, 
Andrei Nomerotski$^{71}$, 
Lyman Page$^{34}$, 
Gonzalo A. Palma$^{72}$, 
Hiranya V. Peiris$^{73,43}$, 
Francesco Piacentini$^{28}$, 
Francesco Piacentni$^{28,74}$, 
Levon Pogosian$^{75}$, 
Chanda Prescod-Weinstein$^{}$, 
Giuseppe Puglisi$^{27,76}$, 
Benjamin Racine$^{25}$, 
Marco Raveri$^{11,20}$, 
Christian L.~Reichardt$^{44}$, 
Mathieu Remazeilles$^{77}$, 
Gra\c{c}a Rocha$^{}$, 
Graziano Rossi$^{78}$, 
John Ruhl$^{79}$, 
Benjamin Saliwanchik$^{69}$, 
Misao Sasaki$^{55}$, 
Emmanuel Schaan$^{16,80}$, 
Alessandro Schillaci$^{21}$, 
Marcel Schmittfull$^{47}$, 
Douglas Scott$^{81}$, 
Neelima Sehgal$^{61}$, 
Leonardo Senatore$^{76}$, 
Huanyuan Shan$^{82}$, 
Blake D.~Sherwin$^{23,24}$, 
Eva Silverstein$^{27}$, 
Sara Simon$^{50}$, 
An\v{z}e Slosar$^{71}$, 
Suzanne Staggs$^{34}$, 
Glenn Starkman$^{79}$, 
Albert Stebbins$^{10}$, 
Radek Stompor$^{30}$, 
Aritoki Suzuki$^{16}$, 
Eric R. Switzer$^{83}$, 
Peter Timbie$^{84}$, 
Matthieu Tristram$^{85}$, 
Mark Trodden$^{86}$, 
Yu-Dai Tsai$^{10}$, 
Caterina Umilt\`a$^{13}$, 
Alexander van Engelen$^{15}$, 
Abigail Vieregg$^{20}$, 
David Wands$^{12}$, 
Yi Wang$^{87}$, 
Nathan Whitehorn$^{88}$, 
W.~L.~K.~Wu$^{11}$, 
Weishuang Xu$^{18}$, 
Matias Zaldarriaga$^{47}$, 
Gong-Bo Zhao$^{89,12}$, 
Yi Zheng$^{90}$, 
Ningfeng Zhu$^{86}$, 
Andrea Zonca$^{91}$\\

\n{\it $^{1}$ \SLAC \\\n$^{2}$ \Oxford \\\n$^{3}$ \LLNL \\\n$^{4}$ \UWC \\\n$^{5}$ \UNM \\\n$^{6}$ \Cornell \\\n$^{7}$ \GRAPPA \\\n$^{8}$ \ITFA \\\n$^{9}$ \JHU \\\n$^{10}$ \FNAL \\\n$^{11}$ \KICP \\\n$^{12}$ \Port \\\n$^{13}$ \Cincinnati \\\n$^{14}$ \ANLHEP \\\n$^{15}$ \CITA \\\n$^{16}$ \LBL \\\n$^{17}$ \PI \\\n$^{18}$ \HarvardPhys \\\n$^{19}$ \SussexAstronomy \\\n$^{20}$ \UChicago \\\n$^{21}$ \Caltech \\\n$^{22}$ \ioa \\\n$^{23}$ \damtp \\\n$^{24}$ \kavli \\\n$^{25}$ \CfA \\\n$^{26}$ \UCI \\\n$^{27}$ \Stanford \\\n$^{28}$ \RomaS \\\n$^{29}$ \INFNRM \\\n$^{30}$ \APC \\\n$^{31}$ \CEADAP \\\n$^{32}$ \JPL \\\n$^{33}$ \BU \\\n$^{34}$ \Princeton \\\n$^{35}$ \UrbanaC \\\n$^{36}$ \ICE \\\n$^{37}$ \IAP \\\n$^{38}$ \IFT \\\n$^{39}$ \UAM \\\n$^{40}$ \UFL \\\n$^{41}$ \JPL \\\n$^{42}$ \Haverford \\\n$^{43}$ \OskarKlein \\\n$^{44}$ \Melbourne \\\n$^{45}$ \UMN \\\n$^{46}$ \Cavendish \\\n$^{47}$ \IAS \\\n$^{48}$ \CCA \\\n$^{49}$ \OSU \\\n$^{50}$ \UMich \\\n$^{51}$ \UTD \\\n$^{52}$ \Columbia \\\n$^{53}$ \YorkU \\\n$^{54}$ \IBS \\\n$^{55}$ \IPMU \\\n$^{56}$ \Buffalo \\\n$^{57}$ \UCD \\\n$^{58}$ \Brown \\\n$^{59}$ \BenGurion \\\n$^{60}$ \INFNFE \\\n$^{61}$ \StonyBrook \\\n$^{62}$ \MIT \\\n$^{63}$ \UW \\\n$^{64}$ \VSI \\\n$^{65}$ \SMU \\\n$^{66}$ \ICTP \\\n$^{67}$ \dunlap \\\n$^{68}$ \NBI \\\n$^{69}$ \Yale \\\n$^{70}$ \UGTO \\\n$^{71}$ \BNL \\\n$^{72}$ \DFI \\\n$^{73}$ \UCL \\\n$^{74}$ \INFNRM \\\n$^{75}$ \SimonFraser \\\n$^{76}$ \KIPAC \\\n$^{77}$ \UoM \\\n$^{78}$ \Sejong \\\n$^{79}$ \CWRU \\\n$^{80}$ \UCBP \\\n$^{81}$ \UBC \\\n$^{82}$ \SHAO \\\n$^{83}$ \GSFC \\\n$^{84}$ \UWMadison \\\n$^{85}$ \ParisSud \\\n$^{86}$ \UPenn \\\n$^{87}$ \HKUST \\\n$^{88}$ \UCLA \\\n$^{89}$ \NAOC \\\n$^{90}$ \KIAS \\\n$^{91}$ \UCSD}

 \pagebreak
 \pagenumbering{arabic}
\section{Executive Summary}
\vspace{-0.08in}
The next generation of instruments designed to measure the polarization of the cosmic microwave background (CMB) will provide a historic opportunity to open the gravitational wave window to the primordial Universe. A dedicated effort over the next decade will lead to a guaranteed detection of the imprint of primordial gravitational waves on the degree-scale {\it B}-mode polarization of the CMB if the predictions of some of the leading models for the origin of the hot big bang are borne out. A detection would reveal a new scale of high energy physics near the energy scale associated with Grand Unified Theories, provide the first evidence for the quantization of gravity, and yield insight into the symmetries of nature, possibly even into deep properties of quantum gravity. For many, a null result would signal a dramatic shift in our understanding of basic early Universe cosmology. CMB data will also provide strong limits on nonlinear processes and phase transitions through constraints on the contribution of gravitational waves to the total energy budget of the Universe.

The CMB provides a unique probe of high energy phenomena through its record of very high temperature, early-Universe physics. The scales explored can be as much as $10^{9}$ times higher than those achieved in terrestrial colliders. Indeed, while the previous decade of work brought about the successful operation of the Large Hadron Collider and discovery of the Higgs boson, the lack of new physics signatures in colliders brings renewed importance to pursuing cosmological probes of high-energy physics. The observational goal that will allow new phenomena to be tested through the CMB is clear: we must measure the polarization to high precision.
 
In the next decade, CMB studies will allow us to address the following key questions: \\
$\bullet$ \hspace{0.1in} How were the seeds for all structure in the Universe created? Is there relic information about their particular quantum origin?\\
$\bullet$ \hspace{0.1in} Did the same phenomenon that laid down the primordial density perturbations also generate primordial gravitational waves? If so, what is their spectrum? What does the signal imply for particle physics at high energies? For gravity? \\
$\bullet$ \hspace{0.1in} Did other highly energetic, nonlinear early-Universe phenomena generate primordial gravitational waves? What do the data tell us about particle physics, including the origin of the hot Universe, phase transitions, and the origin of the matter/anti-matter asymmetry?

The technical groundwork for more sensitive CMB experiments with stronger foreground characterization capability has been laid by the successful satellite and sub-orbital studies of the last decade, including the {\it Planck} satellite \cite{Akrami:2018vks}, and the ground-based ACT \cite{Louis:2016ahn}, SPT \cite{Hanson:2013hsb}, POLARBEAR \cite{Ade:2017uvt}, and BICEP2/Keck \cite{Ade:2015fwj} experiments. To access the target thresholds advocated here, next generation instruments will build on those successes to isolate the cosmic signal from Galactic foregrounds through multi-frequency observations, and reduce the sample variance caused by gravitational lensing either through full-sky observations, or through precise measurements of small-scale fluctuations \cite{Ade:2015zua, Hanson:2013hsb, Ade:2014xna, Ade:2014afa,Madhavacheril:2014slf,Manzotti:2017net}, or both. Through the exciting possibility of a detection as the upper limit on the tensor-to-scalar ratio improved, this decade's instruments catalyzed the theory community to sharpen the understanding of the implications of a {\it B}-mode detection. The stage is now set for the next generation that will either detect primordial {\it B}-modes, or reduce the upper limit on the tensor-to-scalar ratio, $r$, from the current $r\leq 0.06$ at 95\% CL \cite{Ade:2018gkx} by a factor of 10--100 (cf., BICEP Array \cite{Hui:2018cvg}, SPT-3G \cite{Benson:2014qhw}, Simons Observatory \cite{Ade:2018sbj}, CMB-S4 \cite{Abazajian:2016yjj}, and the LiteBIRD \cite{Matsumura:2013aja} and PICO \cite{Sutin:2018onu} satellite concepts). These instruments will achieve, at high confidence, sensitivity thresholds of fundamental importance for our understanding of the origin of the Universe. If vacuum fluctuations during inflation generate primordial gravitational waves, then for $r\gtrsim 0.01$, the inflaton traverses a super-Planckian field range. Models of inflation with characteristic scale near the Planck scale, $M_{\rm Pl}$, and that naturally explain scale-dependence in the density fluctuations by fixing the spectral index to be inversely proportional to the number of e-folds, predict $r\gtrsim 10^{-3}$. If these thresholds are passed without a detection, most textbook models of inflation will be ruled out; and, while the possibility of an early inflationary phase would still remain viable, the data would then force a significant change in our understanding of the primordial Universe.

\section{Sources of gravitational waves in the early Universe}
\vspace{-0.08in}
{\bf A. Vacuum fluctuations from inflation:}
Inflation is an era of accelerated expansion that preceded the current expanding phase, and provided the energy to heat the Universe to a temperature at least high enough for big bang nucleosyntheis (BBN) to occur. As a phenomenological model, inflation successfully reproduces many aspects of the observed Universe, including the lack of spatial curvature, the adiabatic density perturbations, and the super-horizon coherence of perturbations.  During inflation, tiny quantum vacuum fluctuations are amplified and their wavelengths are stretched to cosmological scales by the accelerated expansion. Quantum fluctuations in the matter field(s) during inflation provide the seeds for the growth of structure in the Universe.

\begin{wrapfigure}{r}{0.66\textwidth}
\begin{centering}
\includegraphics[trim=0.4cm 0.2cm 0.3cm 0.8cm, width=0.65\textwidth]{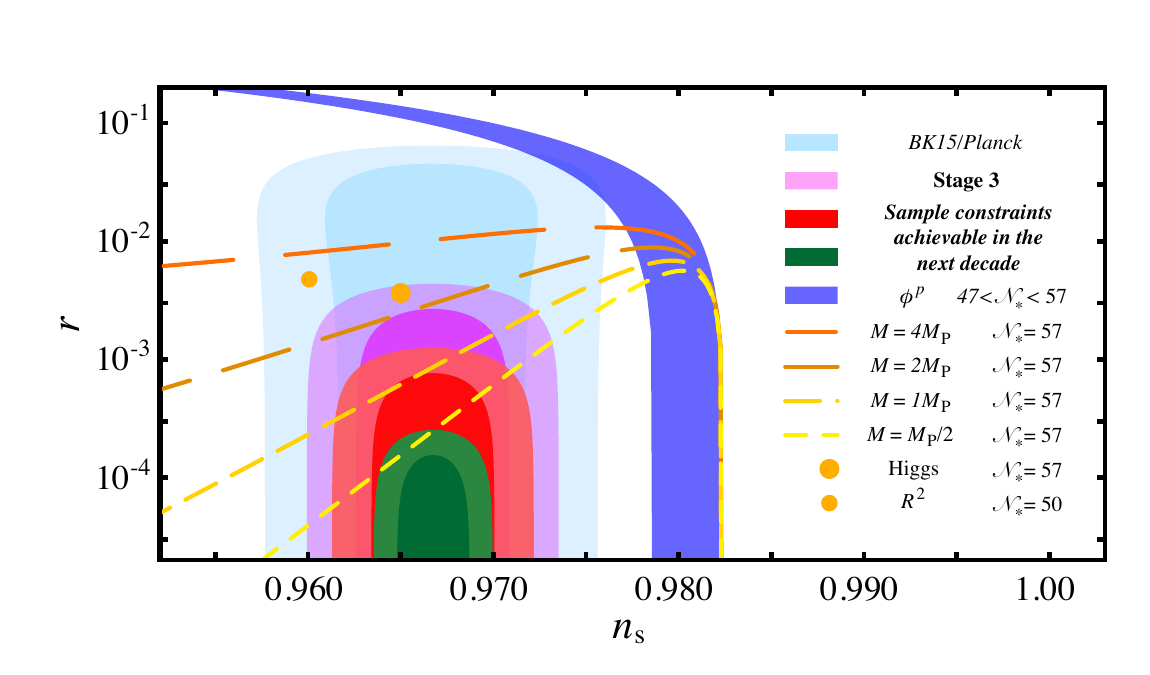}
\caption{
\it{\small Predictions for the tensor-to-scalar ratio $r$ and spectral index $n_{\rm s}$ for some representative single-field inflationary models in which \mbox{$n_{\rm s}-1\propto-1/\mathcal{N}_*$}. This class includes monomial models with $V(\phi)\propto \phi^p$ (dark blue), the Starobinsky ($R^2$) model, and Higgs inflation (orange filled circles). The dashed lines show the predictions of models in this class as function of the scale in the potential. All models with Planckian scale can be detected or excluded in the next decade.}
\label{fig:nsr}}
\end{centering}
\end{wrapfigure}

Quantum fluctuations of spacetime itself produce gravitational waves. Their detection as {\it B}-modes in the polarization of the CMB, with correlations on scales larger than the Hubble scale at the time of last scattering, would provide strong evidence that the tensor fluctuations were produced by the same physics as the observed scalar fluctuations: quantum fluctuations of the vacuum. 


The amplitude of tensor fluctuations relative to the scalar fluctuations, $r$, directly reveals the energy scale of inflation. A measurement of $r$ fixes the dominant component of the energy density during inflation, $V$, via the relation 
\begin{equation}
V^{1/4}=1.04\times 10^{16} \; {\rm GeV}\left(\frac{r}{0.01}\right)^{1/4}\,.
\label{eq:Infscale}
\end{equation}
The enormously high energy scale in the prefactor suggests that measurements of the primordial fluctuations might shed light on high energy physics in a regime far beyond the standard model of particle physics. Ideally the particle content during the inflationary era can be constrained and eventually connected to other data across the vast range of scales separating inflation from laboratory particle physics.


While aspects of the scalar perturbations (e.g., the shape of the power spectrum and non-Gaussianity) have well-explored connections to inflationary particle content, the relative amplitude of the {\it B}-mode signal goes further and can provide a unique probe of quantum gravity through its relation to the inflaton field range. When a scalar field sources inflation, the distance $\Delta\phi$ that it moves in field space during inflation is related to the tensor-to-scalar ratio by~\cite{Lyth:1996im}
\begin{equation}
    \frac{\Delta\phi}{M_{\rm Pl}}\gtrsim\left(\frac{r}{8}\right)^{1/2} \mathcal{N}_*\gtrsim\left(\frac{r}{0.01}\right)^{1/2}\,,
    \label{eq:lyth}
\end{equation}
where $M_{\rm Pl}$ is the Planck mass, and $\mathcal{N}_*$ is the number of e-folds between the end of inflation and the time when the pivot mode, $k_*=0.05$ Mpc$^{-1}$, exits the Hubble volume during inflation. The right-hand side of Eq.~(\ref{eq:lyth}) uses a conservative lower limit $\mathcal{N}_*\gtrsim 30$ \cite{Liddle:2003as,Baumann:2006cd,Easther:2006qu}. 
Why is this field range of interest? It is generally expected that {\it any} theory of quantum gravity will introduce new degrees of freedom at or below the Planck scale that can interact with the inflaton. Unless a symmetry forbids these interactions, one expects sub-Planckian features in the inflaton potential that prevent $\Delta\phi\gtrsim M_{\rm Pl}$. The relation in Eq.~(\ref{eq:lyth}) then implies that a detection of $r\gtrsim0.01$ would be strong evidence for such a symmetry. The importance of a detection of $r$ may be even more profound: the permissibility of a Planckian field range in a consistent, nonperturbative theory of quantum gravity, irrespective of the symmetries that might protect the inflaton potential, remains the subject of considerable debate~\cite{ArkaniHamed:2006dz,Ooguri:2006in,Ooguri:2018wrx}. It is truly remarkable that CMB data can weigh in on quantum gravity. Searching for evidence of large-field models sets the threshold for {\bf Science Objective A1} in Table \ref{tab:goals} of the Summary section.
\begin{wrapfigure}{r}{0.55\textwidth}
\begin{centering}
\includegraphics[trim=0.2cm 0.2cm 0.15cm 0.2cm,width=0.55\textwidth]{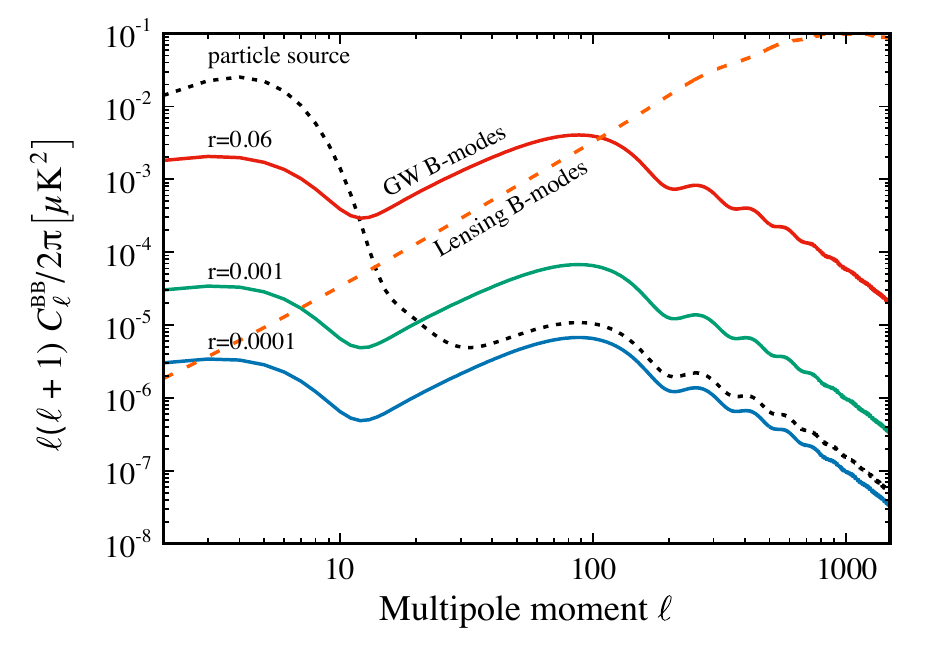}
\caption{\it{\small {\it B}-mode spectrum from vacuum fluctuations (solid lines, top to bottom: \mbox{$r = 0.06, 10^{-3}, 10^{-4}$)}, from gravitational lensing of E-modes (dashed orange line), and from an example with contribution from vacuum fluctuations with \mbox{$r=1.6\times10^{-4}$} and particle production, generating strong large angular scale {\it B}-modes (dotted line)~\cite{Namba:2015gja}).}
\label{fig:ClBB}}
\end{centering}
\vspace{-0.1in}
\end{wrapfigure}
Following on the discussion above, one way to classify inflation models is by the typical scale of structure in the potential - the scale over which its value changes appreciably. Improved constraints on $r$, together with better characterization of the scalar spectrum (via the spectral index $n_{\rm s}$), will allow large classes of single-field inflation models to be ruled out in the next decade, even in the absence of a detection of primordial gravitational waves. A class of models of particular interest are those that explain the observed value of the spectral index via $n_{\rm s}-1\propto-\frac{1}{\mathcal{N}_{*}}$~\cite{Mukhanov:2013tua,Roest:2013fha,Creminelli:2014nqa,Abazajian:2016yjj}. An upper limit of $r< 10^{-3}$ would rule out all such models that naturally, in this sense, explain the current central value of the spectral index, and that have a characteristic scale set by the Planck scale $M_{\rm Pl}$. Models that would be excluded include the historic ``$R^2$" model~\cite{Starobinsky:1980te} (predicting $r> 0.003$) and monomial inflation, $V(\phi)\propto \phi^{p}$, including string theory motivated models \cite{Silverstein:2008sg}.
Finding evidence of natural, Planck-scale models, or ruling them out, sets {\bf Science Objective A2} and is illustrated in Figure~\ref{fig:nsr}. 

Regardless of the details of the inflationary model, the gravitational-wave background should have super-horizon coherence~\cite{Baumann:2009mq,Lee:2014cya}.
Confirming this prediction through measurements of correlations in the {\it B}-mode signal on angular scales $\theta\gtrsim 2^\circ$ is {\bf Science Objective A3}.

{\bf B. Particle sources during inflation:} During inflation, additional fields (besides the inflaton) can act as extra sources of gravitational waves that can dominate over the vacuum fluctuations. The presence of those fields in the inflationary context is natural from a top-down perspective; string theory, for example, provides plenty of candidates \cite{Baumann:2014nda}. In the well-studied case of axion inflation~\cite{Pajer:2013fsa}, the sourced tensor power spectrum can deliver a large (and chiral) gravitational wave signal even if the scale of inflation is well below the scale in Eq.~(\ref{eq:Infscale}). In general, the phenomenology of sourced gravitational waves is strikingly different from that of amplified vacuum fluctuations. Figure \ref{fig:ClBB} compares a representative particle-sourced {\it B}-mode power spectrum to that from vacuum fluctuations. Although the model must be tuned to give a large signal at measurable scales, it provides an important example of a distinguishable spectrum.

Quite generally, if primordial {\it B}-modes are detected, distinguishing models for their origin will require a careful analysis of the spectrum. A sourced tensor spectrum can present broad features~\cite{Namba:2015gja,Dimastrogiovanni:2016fuu,Thorne:2017jft}, characterized by a detectably large (and running) spectral index $n_{\rm t}\equiv\frac{d\ln P_{\rm t}}{d\ln k}={\cal O}(1)$ (Fig. \ref{fig:ClBB}, {\bf Science Objective B1}). The sourced spectrum is chiral, which can be seen in non-zero $EB$ and $TB$ correlations \cite{Lue:1998mq}, and a fully chiral tensor spectrum can be detectable at the $2 \sigma$ level in a cosmic variance-limited experiment as long as $r> 0.01$~\cite{Gluscevic:2010vv,nati_2017}. Finally, sourced {\it B}-modes may have detectably large non-Gaussianities spanning a rich class of bispectra \cite{Shiraishi:2016yun}. 

{\bf C. A gravitational wave probe of post-inflationary particle physics:\flushright}
\begin{wrapfigure}{r}{0.63\textwidth}
\begin{centering}
\includegraphics[trim=0.2cm 0.2cm 0cm 0.4cm,width=0.63\textwidth]{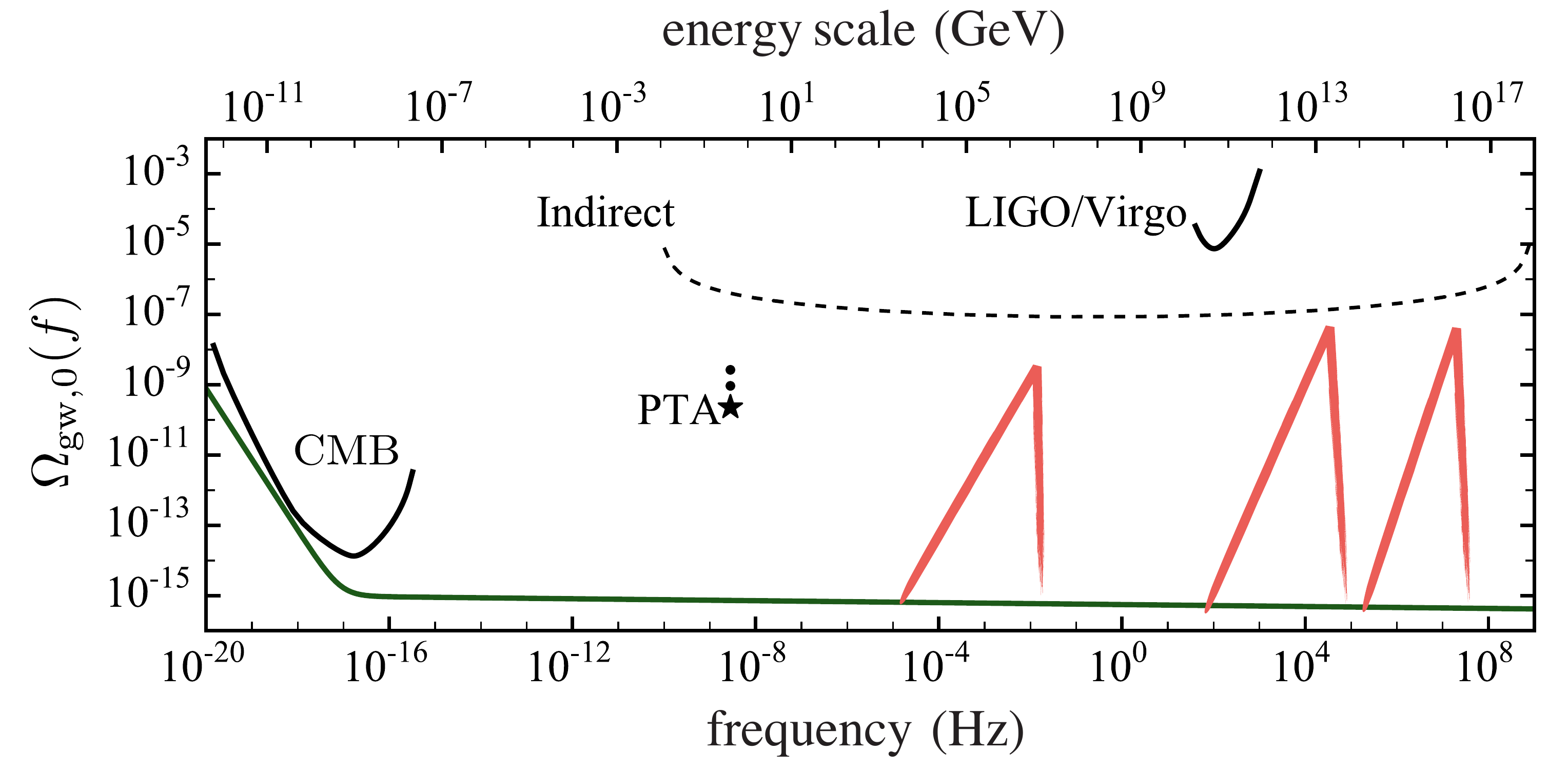}
\caption{\it{\small CMB constraints on the stochastic gravitational wave background. The solid black line is the 2015 direct CMB limit of $r<0.12$ ($n_{\rm t}=0$) on the low-frequency stochastic background. The indirect constraint (dashed line) comes from the $\Omega_{gw}$ limit from the CMB together with other cosmological data (\cite{Lasky:2015lej}, 2015 data). The primordial spectrum is shown for $r=0.11$ and $n_{\rm t}=-r/8$ (green). Curves in red show schematic example signals that could arise from (left) the electromagnetic phase transition \cite{Caprini:2015zlo}, (center) a new physics phase transition, and (right) re-heating. }
\label{fig:OmegaGW}}
\end{centering}
\end{wrapfigure}
\vspace{-0.1in}
In a cooling universe one may expect short periods with complex, nonlinear field dynamics, including phase transitions, non-perturbative particle production, and the formation of solitons/defects (see, e.g., \cite{Amin:2014eta}). These processes generically produce a gravitational-wave spectrum extending over just a few decades in frequency. The signal is sharply peaked, with the peak frequency scaling as $f \sim \sqrt{H_0 H_*}$ where $H_*$ is the Hubble scale at the time of production \cite{Easther:2006gt,Dufaux:2007pt}. For example, the reheating process after inflation and phase transitions associated with the origin of the matter/anti-matter asymmetry may lead to spectra of this type, shown in \mbox{Figure \ref{fig:OmegaGW}}. 

While a few narrow frequency bands are being (or will be) probed by direct detection (e.g., LIGO at $f\sim 10^{2}\rm Hz$, LISA at $f\sim 10^{-3}\rm Hz$ \cite{Caprini:2015zlo}), the CMB is sensitive to the \emph{total} integrated number of relativistic degrees of freedom at the time of recombination \cite{Maggiore:1999vm} and constrains the total energy density in gravitational waves integrated over all sub-horizon wavelengths. Assuming no radiation-like new particles, current data restrict $\Omega_{\rm gw,0} \lesssim 1$--$2 \times 10^{-6}$. Limits on the integrated $\Omega_{\rm gw,0}$ from the CMB constrain the strength of all nonlinear particle processes in the early Universe, and can be combined with data from direct-detection experiments to constrain the shape of the primordial spectrum \cite{Lasky:2015lej}. {\bf Science Objective C1} is the expected improved upper limit on $\Omega_{\rm gw,0}$.

{\bf D. Non-inflationary models for the origin of the cosmological perturbations:}
Inflation is not the only proposed mechanism for generating the cosmological perturbations. Alternative scenarios include Ekpyrotic and Cyclic models \cite{Steinhardt:2001vw,Khoury:2001wf}, String Gas Cosmology \cite{Brandenberger:1988aj,Battefeld:2005av}, and a matter bounce \cite{Wands:1998yp,Finelli:2001sr}. Much less community effort has gone into developing these alternatives, but the expectation is that the original Ekpyrotic scenario predicts a very small value of $r$ \cite{Boyle:2003km}, while String Gas Cosmology predicts a value observable with near-term experiments. In many realizations of non-inflationary scenarios, the predictions are model dependent and an observation of $r$ alone may not be enough to distinguish them from inflation. However, when the amplitude of $r$ is combined with constraints from other observables (such as the shape of the gravitational wave spectrum) it may be possible to distinguish these scenarios from inflation \cite{Brandenberger:2016vhg}.
\vspace{-0.15in}

\section{Summary and Recommendations}
\vspace{-0.08in}

CMB polarization experiments will provide a unique observational window into fundamental particle physics and gravity above about 10 TeV, through searches for the signatures of gravitational waves. The sentiment expressed in the last decadal review still stands: in pushing forward our understanding of the earliest phases of the Universe, ``The most exciting quest of all is to hunt for evidence of gravitational waves that are the product of inflation itself." \cite{NAP12951}. At the same time, new data may uncover signatures beyond the simplest predicted spectrum, either from non-vacuum sources during inflation, from energetic particle processes after inflation, or from alternatives to inflation. With either a detection or an upper limit, next generation polarization surveys have the potential to continue the rich legacy of cosmological discovery as the driver of innovation in fundamental particle physics and gravity.

\begin{table}[h!]
    \centering
    \begin{tabular}{l|c}
    \hline\hline
       
      {\bf Gravitational wave physics} & {\bf Target} \\
      
       \hline

        {\bf A1}: Large-field inflation models$^*$ & $r\gtrsim 0.01$   \\
        {\bf A2}: Natural, Planck-scale inflationary potentials $^*$ &$r\gtrsim 0.001$ \\
        {\bf A3}: Primordial origin for {\it B}-modes&  correlations on scales $\theta\gtrsim2^\circ$ \\
       \hline
       {\bf B1}: Matter sources& $\sigma(n_{\rm t})>1$  \\  
        \hline
        {\bf C1}: Energetic phase transitions &$\Omega_{\rm gw}< 10^{-7}$  \\
        \hline
        \end{tabular}\\[.2cm]
        \hskip -2.25cm Current bounds: \hskip .5cm$r<0.06$ at 95\% CL~\cite{Ade:2018gkx}\quad\text{and}\quad $\Omega_{\rm gw}< 10^{-6}$~\cite{Lasky:2015lej}.
    
    \caption{Science objectives, where a detection above the target limit would provide evidence for the listed phenomena. The asterisk indicates statements that rely on single-field (adiabatic) inflationary vacuum fluctuations as the main gravitational wave source. 
    \label{tab:goals}}

\end{table}
\pagebreak

\end{document}